\documentclass[referee]{raa}
\usepackage{amssymb}
\usepackage{graphicx,times}

\begin{document}

\title {Modelling the WMAP large-angle anomalies as an effect of a
local density inhomogeneity}
\author{Liping He\inst{1,2}, Quan Guo\inst{1}}
\institute{National Astronomical Observatories,
Chinese Academy of Sciences, Beijing 100012, China \\
\email{hlp@bao.ac.cn} \and
 Graduate School of Chinese Academy of Sciences, Beijing 100049,
China\\
 }
\abstract{We investigate large-angle scale temperature anisotropy in
the Cosmic Microwave Background\,(CMB) with the Wilkinson Microwave
Anisotropy Probe\,(WMAP) data and model the large-angle anomalies as
the effect of the CMB quadrupole anisotropies caused by the local
density inhomogeneities. The quadrupole caused by the local density
inhomogeneities is different from the special relativity kinematic
quadrupole. If the observer inhabits a strong inhomogeneous region,
the  local quadrupole should not be neglected. We calculate such
local quadrupole under the assumption that there is a huge density
fluctuation field in direction $(284^{\circ},74^{\circ})$, where the
density fluctuation is $10^{-3}$, and its center is $\sim
112h^{-1}\,\rm {Mpc}$ away from us. After removing such mock signals
from WMAP data, the power in quadrupole, $C_2$, increases from the
range $(200\sim260\mu \rm{K^2})$ to $\sim1000\mu \rm{K^2}$. The
quantity S, which is used to estimate the alignment between the
quadrupole and the octopole, decreases from $(0.7\sim0.74)$ to
$(0.31\sim0.37)$, while the model predict that $C_2=1071.5\mu
\rm{K^2}$, $S=0.412$. So our local density inhomogeneity model can,
in part, explain the WMAP low-$\ell$ anomalies.
 }

\authorrunning{Liping He, Quan Guo}
\titlerunning{Modelling the WMAP low-$\ell$ anomalies as an effect of a local
density inhomogeneity}
 \maketitle

\section{Introduction}
Although the WMAP data are regarded as a dramatic confirmation of
standard inflationary cosmology (Vale~\cite{val05}; de
Oliveira-Costa \& Tegmark~\cite{oli06}; Gazta\~{n}aga et
al.~\cite{gaz03}; Gordon et al.~\cite{gor05}), some anomalous
features have emerged (Inoue \& Silk~\cite{ino06}; Campanelli et
al.~\cite{cam06}; Dominik et al.~\cite{dom04}). Firstly, the
amplitude of the quadrupole is substantially less than the
expectation from the best-fit $\Lambda \rm{CDM}$ standard model
(Abramo et al.~\cite{abr06}; de Oliveira-Costa et al.~\cite{oli04};
Efstathiou~\cite{efs04}), which was found by COBE a decade ago
(Bennett et al.~\cite{ben96}) and confirmed by WMAP (Spergel et
al.~\cite{spe03}). Secondly, the quadrupole and octopole indicate an
unexpectedly high degree of alignment (Spergel et al.~\cite{spe03};
de Oliveira-Costa et al.~\cite{oli04},~\cite{oli06};
Schwarz~\cite{sch04}; Land et al.~\cite{lan05}; Hansen et
al.~\cite{han04a}a; Eriksen et al.~\cite{eri04}).

Recently, many efforts have been devoted to explain the origin of
the anomalies. They can be systematic error, statistical flukes,
improper subtraction of known foreground, or an unexpected
foreground (Copi et al.~\cite{cop04}, ~\cite{cop05},~\cite{cop06}).
The WMAP team claims that there are no unexpected systematic errors
(Bennett et al.~\cite {ben03}; Finkbeiner~\cite{fin04}), and Copi et
al.~(\cite{cop04}, \cite{cop05}, \cite{cop06}) noted that the
anomalies are most unlikely to be due to residual foreground
contamination. Several authors attempted to explain the anomalies in
terms of a new foreground (Abramo et al.~\cite{abr06}; Gordon et
al.~\cite{gor05}; Bennett et al.~\cite {ben03};
Finkbeiner~\cite{fin04}; Prunet et al.~\cite {pru05}; Rakic et
al.~\cite{rak06}).

Abramo et al.~(\cite{abr06}) showed circumstantial evidences that an
extended foreground near the dipole axis could distort the CMB. They
proposed that the possible physical mechanism, which can produce
such a foreground, is the thermal Sunyaev-Zeldovich\,(SZ) effect.
But the SZ model, as presented by them, cannot account for the
anomalous quadrupole and octopole successfully. Therefore, they
thought that the Ress-Sciama \,(RS) effect (Rakic et
al.~\cite{rak06}), or the combination of SZ effect and RS effect may
be responsible for the foreground. Many other authors suggested that
the large-angle anomalies are affected by local inhomogeneities
(Tomita~\cite{tom05a}a, ~\cite{tom05b}b; Vale~\cite{val05}).
However, when they applied a model in which the Local Group is
falling into the center of the Shapley supercluster, the discrepancy
between the observed data and the model prediction became even
worse. (Rakic et al.~\cite{rak06}; Inoue et al.~\cite{ino07}).

Inoue et al. (2007) explored the large angular scale temperature
anisotropies due to homogeneous local dust-filled voids in a flat
Friedmann-Robertson-Walker universe. They found that a pair of voids
with radius $(2\sim3)\times10^2\rm{h^{-1}Mpc}$ and density contrast
$\delta_m\sim0.3$ might help explain the observed large-angle CMB
anomalies. While Wu \& Fang~(\cite{wu94}) explored the possibility
that the CMB is affected by local density inhomogeneities basing on
Tolman-Bondi model. They calculated the quadrupole amplitude of the
local collapse model with the general relativity (GR). The results
show that the CMB anisotropies from the local quadrupole
contribution can be different from the special relativity (SR)
kinematic quadrupole by a factor as large as 3, which depends on the
size and density fluctuation of the region the observer inhabits.
Therefore, if we live in a large density fluctuation area, the local
quadrupole might be significant in the CMB observations.

The goal of this paper is to examine whether such local quadruple
could account for the observed large-angle CMB anomalies in WMAP
data. Our analysis is based on the 1-year, 3-year and 5-year WMAP
Internal Linear Combination maps (Spergel et al.~\cite{spe06};
Hinshaw et al.~\cite{hin06}) (henceforth ILC1, ILC3 and ILC5). We
try to remove the mock CMB foreground caused by the effect described
in Wu \& Fang~(\cite{wu94}) for each observed CMB map under the
assumption that we are in a huge density fluctuation area. The
parameters of the area we adopted are based on Kocevski \&
Ebeling~(\cite{koc06}) and Watkins et al.~(\cite{wat08})'s work. We
reanalyze the WMAP data by using the multipole vector framework in
Section 2. In Section 3, we review the estimate of the foreground of
Wu \& Fang~(\cite{wu94}) and present the result of our examination.
We conclude in Section 4.

\section{Large-angle anomalies of CMB}
In this section, we re-investigate the anomalies reported from the
WMAP maps on very large angular scale. As we already remarked, the
angular power in quadrupole, $C_2$, is less than expected. To
measure $C_2$, we expand the temperature anisotropy in terms of
spherical harmonics (Campanelli et al.~\cite{cam06}; Copi et
al.~\cite{cop04})
\begin{equation}
\Delta
T(\theta,\phi)=\sum_{\ell=1}^{\infty}\sum_{m=-\ell}^{\ell}a_{\ell
m}Y_{\ell m}(\theta,\phi).
\end{equation}
And the angular power spectrum is defined as
\begin{equation}
C_{\ell}\equiv \frac{1}{2\ell+1}\sum_{m=-\ell}^{\ell}|a_{\ell m}|^2.
\end{equation}

A simple way to quantify the peculiar alignment of the quadrupole
and octopole is to use the multipole vectors. In the multipole
vector representation, the $\ell-th$ multipole of the CMB,
$T_{\ell}$, can be written in terms of a scalar $A^{(\ell)}$ and
$\ell$ unit vectors $\{\hat{\upsilon}^{(\ell,i)}|i=1,\cdots,\ell\}$
(Dominik et al.~\cite{dom04}; Copi et al.~\cite{cop05},
~\cite{cop06})
\begin{equation}
T_{\ell}\approx
A^{(\ell)}\prod_{i=1}^{\ell}(\hat{\upsilon}^{(\ell,i)}\cdot\hat{e}).
\end{equation}

For the statistical comparison, we use the area vectors
\begin{equation}
\vec{w}^{(\ell;i,j)}\equiv
\hat{\upsilon}^{(\ell,i)}\times\hat{\upsilon}^{(\ell,j)}.
\end{equation}

The alignments between the quadrupole area vector and the three
octopole area vectors can be evaluated by the magnitudes of the dot
products between $\vec{w}^{(2;1,2)}$ and each $\vec{w}^{(3;i,j)}$
\begin{eqnarray}
A_1\equiv |\vec{w}^{(2;1,2)} \cdot \vec{w}^{(3;1,2)}|
\nonumber\\
A_2\equiv |\vec{w}^{(2;1,2)} \cdot \vec{w}^{(3;2,3)}|\\
A_3\equiv |\vec{w}^{(2;1,2)} \cdot \vec{w}^{(3;3,1)}| \nonumber.
\end{eqnarray}

The widely used estimator that checks for alignments of the
quadrupole and octopole planes is the average of the dot products
(Dominik et al.~\cite{dom04}; Abramo et al.~\cite{abr06}; Katz \&
Weeks~\cite {kat04}; Schwarz~\cite{sch04})
\begin{equation}
S=\frac{1}{3}\sum_{i=1}^{3}A_i.
\end{equation}

Given a CMB map, the harmonic components can be easily extracted
with the HEALPix\footnote{http://healpix.jpl.nasa.gov/} (G\'{o}rski
et al.~\cite{gos05}) software, and the multipole vectors can be
calculated by the code provided by Copi et al.~(\cite{cop04}). Our
analysis is based on the 1-year, 3-year and 5-year WMAP full sky
maps (ILC1, ILC3, ILC5). The values of $C_2$ and S for ILC1, ILC3
and ILC5 are listed in Table 1. We can see that $C_2$ lies in the
range $(200\sim260\mu\rm K^2)$.

In order to compare with the $\Lambda \rm{CDM}$ standard model,
$10^6$ mimic CMB maps are generated with Monte Carlos\,(MC)
simulation based on theoretical CMB power spectrum predicted by
$\Lambda \rm{CDM}$ model, which is generated by
CAMB\footnote{http://camb.info/} (Lewis et al.~\cite{lew00}) package
with the best-fitting cosmological parameters estimated from WMAP
(Hinshaw et al.~\cite{hin09}). In the $\Lambda \rm{CDM}$ model, the
power in quadrupole is $C_2=1071.5\,\mu \rm{\rm{K^2}}$, while the
power in quadrupole for the ILC1, ILC3 and ILC5 are $C_2=204.4\,\mu
\rm{\rm{K^2}}$, $C_2=260.3\,\mu \rm{\rm{K^2}}$, $C_2=254.1\,\mu
\rm{\rm{K^2}}$. Clearly, the WMAP data have a low power in
quadrupole compared to $\Lambda \rm{CDM}$ model.

Fig 1 is the histogram of the S statistics generated from $10^6$
Gaussian random, statistically isotropic MC mock maps. The average
value of S from $10^6$ MC simulations is $S_{\Lambda
\rm{CDM}}=0.412$, which is much lower than the S statistics from
WMAP data, that is $S=0.744$ for ILC1, $S=0.700$ for ILC3, and
$S=0.726$ for ILC5. The final rank in table 1 lists the odds
$P(S_{\Lambda \rm{CDM}}>S)$ of finding a value among the $10^6$ MC
maps larger than the one observed, from which one can see that the
probabilities are $0.8\%$ for ILC1, $2.1\%$ for ILC3, and $1.2\%$
for ILC5. This means that the alignment between quadrupole and
octopole for each WMAP map is significant.

These alignments could be explained by an unexpected foreground
caused by a local collapse due to the second-order effect of the
density fluctuation area (Wu \& Fang~\cite{wu94}). In next section,
we will briefly discuss this foreground.

\section{Hypothetical foreground induced by super large structure}
The CMB temperature anisotropy produced by a locally spherical
collapse can be modeled basing on a Tolman-Bondi universe solution
(Wu \& Fang~\cite{wu94}).

Because we are interested in the effect of a local density
fluctuation, in the following we only consider the case of
$X_0<X_c$, where $X_0=x_0/t_e$, $x_0$ is the distance between the
observer and the center of the perturbation, $X_c=x_c/t_e$, $x_c$ is
the size of the perturbed region. When the initial density
perturbation $\delta_0$ is assumed to be constant in the region
$x\leq x_c$, the first-order solution consists mainly of two parts:
a monopole term and a dipole term which we are familiar with. The
second-order solution of $\Delta \textsf{T}/\textsf{T}$ is (Wu \&
Fang~\cite{wu94})
\begin{equation}
\frac{\Delta\textsf{T}}{\textsf{T}}=\delta_0^2 \bigg[ \left(
\frac{3}{175}X_c^2-\frac{11}{1575}X_0^2\right)T_0^{2/3}+\frac{4}{175}T_0X_0\cos
\Psi+\frac{2}{225}T_0^{2/3}X_0^2\cos 2\Psi \bigg],
\end{equation}
where $T_0=(1+z_d)^{3/2}$ and $z_d$ is the redshift at decoupling
time $t_e$, and $\Psi$ is the incidence angle of the photon.

When the terms of the order of $\delta_0^2$ and $T_0^{1/3}$ are
taken into account, the quadrupole anisotropy caused by local
density fluctuation should be (Wu \& Fang~\cite{wu94})
\begin{eqnarray}
\left(\frac{\Delta
\textsf{T}}{\textsf{T}}\right)_q=\frac{2}{225}T_0^{2/3}X_0^2\delta_0^2\cos
2\Psi+T_0^{1/3}X_0^2\Delta_q\delta_0^2
\end{eqnarray}
 and
\begin{eqnarray}
\Delta_q & = &
-\frac{19X_c}{3780}-\frac{1}{X_0}\left(\frac{X_0}{140X_c}+\frac{229X_0^3}
{61440X_c^3}+\frac{261X_0^5}{81920X_c^5}+\frac{3X_0^7}{4096X_c^7}\right)
{}\nonumber\\
& & {}+X_0\left[\frac{41X_0}{9800X_c}-\frac{1333X_0^3}{2064384X_c^3}
+\frac{467X_0^5}{5734400X_c^5}+\frac{3833X_0^7}{11468800X_c^7}+O\left(\frac{X_0^9}{X_c^9}\right)
\right]{}.
\end{eqnarray}

The first term in the left-hand side of equation (8) is the SR
kinematic quadrupole anisotropy. Equation (8) tells us that if
higher orders are involved, the SR kinematic quadrupole may not
always be a good approximation of the quadrupole produced by a local
collapse. The local quadrupole anisotropy strongly dependents on the
size, matter density in the peculiar field, and the position of the
observer. Fig 2 shows the quadrupole amplitude as a function of the
distance between the observer and local gravitational field $x_{0}$.
The SR kinematic quadrupole is denoted by solid curve, and the local
quadrupole is denoted by dotted curve. We assume $x_c=1000
h^{-1}\,\rm{Mpc}$ to satisfy $x_c>x_0$. The quadrupole showing in
Fig. 2 is along the center of the perturbation. Fig 3 shows the
relationship between the amplitude of local quadrupole and the
radius of the local gravitational field $x_{c}$ for $x_0=112
h^{-1}\,\rm{Mpc}$. $x_c$ changes from $150 h^{-1}\,\rm{Mpc}$ to
$1000 h^{-1}\,\rm{Mpc}$. Because the distance of the observer to the
center of the collapse should at least be greater than the distance
to the Great Attractor, which is estimated to be $80
h_{50}^{-1}\,\rm{Mpc}$. Therefore, it would be reasonable to take
the lower value of $x_c=150 h^{-1}\,\rm{Mpc}$ which is about 2 times
of the distance to the Great Attractor and the higher value of
$x_c=1000 h^{-1}\,\rm{Mpc}$ which is about the size of horizon (Wu
\& Fang~\cite{wu94}). We find that the influence of $x_{0}$ on the
amplitude of local quadrupole is about one magnitude larger than the
influence of $x_{c}$. When $x_0$ is fixed, the results change little
with $x_{c}$. Fig 4 shows the corrected $C_2$ of ILC5 as a function
of $x_c$ when $x_0=112 h^{-1}\,\rm{Mpc}$. It turns out that
$C_2=1022.3 \rm{\mu K}^2$ for all values of $x_c$.

In order to explain the large-angle anomalies we propose a model
that we are in a large density fluctuation area. As Kocevski \&
Ebeling~(\cite{koc06}) suggests that $56\%$ of the Local
Group's\,(LG) peculiar velocity is induced by more distant
overdensities between 130 and 180\,$\rm{Mpc}$ away. Watkins et
al.~(\cite{wat08}) also notes that the bulk flow within a Gaussian
window of radius $50\,\rm{Mpc}$ is $407\pm81\,\rm{km\, s^{-1}}$
toward $l=287^\circ\pm9^\circ$, $b=8^\circ\pm6^\circ$, and roughly
$50\%$ of the LG's motion is due to sources at greater depths.
Interestingly, we find that a region with a density fluctuation
$\delta\sim 10^{-3}$ over a distance $\sim112\,h^{-1} \rm{Mpc}$ away
on the direction of $(284^{\circ},74^{\circ})$ may be responsible
for the origin of the anomalies on large angular scales. We compute
the mock foreground (equation (8)) using these parameters. Fig 5
shows the map of the contribution of CMB anisotropies caused by the
local density fluctuation.

After subtracting such a mock foreground from the CMB sky maps of
the WMAP observation, we find that the power in quadrupole will
dramatically increase and the alignment of the quadrupole and
octopole plane will be weakened. In Table 2 we compare the
quadrupole and S obtained from the "foreground-corrected" WMAP data
to those obtained from fiducial $\Lambda\rm{CDM}$ model. The powers
in quadrupole of the three WMAP maps increase to $C_2=1064.2\,\mu
\rm{K^2}$, $C_2=1034.2\,\mu \rm{K^2}$, $C_2=1022.3\,\mu \rm{K^2}$,
respectively, which is apparently in much better agreement with the
$\Lambda \rm{CDM}$ model. Furthermore, from the S statistics, one
can see that the frequencies $P(S_{\Lambda \rm{CDM}}>S)$ of finding
a $\Lambda \rm{CDM}$ simulation with a S value larger than that from
WMAP seem to converge to $75.1\%$ for ILC1, $61.5\%$ for ILC3,
$62.2\%$ for ILC5. Therefore, if such a large scale structure exist,
the foreground model presented here can not be neglected.

We evaluate the probability that the primary quadrupole is cancelled
by the local quadrupole. We generate $2000$ CMB maps, which have
random quadrupole orientations, with the HEALPix software, and the
input theoretical power spectra, $C(\ell)$, are generated by the
CAMB package. Then we combine the foreground with the random,
statistically isotropic CMB maps. We find that about $\sim 28\%$ of
the quadrupole are consistent with the observed WMAP five year
values, that is
$C_2=223.479\pm978.367$\footnote{http://lambda.gsfc.nasa.gov/product/map/}.
Therefore, our model can explain part of the anomalies. But the
large errorbar in the quadrupole measurement may also be responsible
for the large number $28\%$.

\section{Conclusions}
In this paper, we have re-investigated the anomalies in  WMAP data.
The power in quadrupole is found to be $C_2=204.4\,\mu \rm{K^2}$ for
ILC1, $C_2=260.3\,\mu \rm{K^2}$ for ILC3 and $C_2=254.1\,\mu
\rm{K^2}$ for ILC5, while the power in quadrupole for the standard
$\Lambda \rm{CDM}$ model is $C_2=1071.5\,\mu \rm{K^2}$. It is
obvious that the power in quadrupole is less than the expected. By
comparing the distribution of the S statistics from WMAP data to
those from $10^6$ MC simulation mimic CMB maps, we found that they
are consistent at the level of $0.8\%$ for ILC1, $2.1\%$ for ILC3
and $1.2\%$ for ILC5. These results indicate that the quadrupole and
octopole planes are aligned strongly.

We provide a possible explanation for the anomalies in WMAP data by
using the foreground model caused by a large density fluctuation.
The model depends on the matter distribution, and the position of
the observer. So we assumed that there is a large-scale structure in
direction $(284^{\circ},74^{\circ})$, the center is $\sim
112h^{-1}\,\rm{Mpc}$ away from us, and the density fluctuation is
$10^{-3}$. After subtracting the mock foreground caused by such area
from the WMAP data ILC1, ILC3 and ILC5, we found that the power in
quadrupole, $C_2$, increases to $(\sim 1000\,\mu \rm{K^2})$ level,
and the S decreases to $0.31\sim 0.37$ level, which agrees with the
prediction from the standard $\Lambda \rm{CDM}$ model. To conclude,
the local gravitational collapse might be responsible for explaining
the origin of the large-angle CMB anisotropy.

Recently, it has been suggested by many researchers that the local
inhomogeneities can account for the large angular scales anomalies
(Tomita~\cite{tom05a}a, ~\cite{tom05b}b; Vale~\cite{val05}).
However, none of the proposed models can successfully explain the
anomalies (Inoue \& Silk~\cite{ino06}). Because it is well known
from the GR that in a linear approximation, the behavior of a
comoving object in an expansion or collapsing metric can not be
equivalently described as SR Doppler motion if the higher orders are
involved. The amplitude of the kinematic quadrupole is about $13\%$
of the cosmic quadrupole (Wu \& Fang~\cite{wu94}). Therefore the CMB
quadrupole anisotropy calculated as an effect of a local density
inhomogeneity can not be approximated by a SR effect, which is the
main reason why we have derived different results from others.

However, many other specific features of the anomalies have been
discovered, such as anomalously cold spots on angular scales
$\sim10^\circ$ (Vielva et al.~\cite{vie04}; Cruz et
al.~\cite{cru05}), and asymmetry in the large-angle power between
opposite hemispheres(Eriksen et al.~\cite{eri04}; Hansen et
al.,~\cite{han04b}b; Sakai \& Inoue ~\cite{sak08}). We have not
interpreted these anomalies with our model explicitly, so further
research is expected.

\section*{Acknowledgement}
We would like to thank Wen Xu, Huan-Yuan Shan, Xiao-Chun Mao,
Xin-Juan Yang, Nan Li and Qian Zheng for helpful comments and
discussions. And we are grateful to WMAP team for providing such a
superb data set. We also thank Wen Xu for careful reading on the
draft manuscript.

\begin{table}[!hbp]
\begin{center}
\caption[]{Power in quadrupole $C_2$ and alignments of the CMB maps
for ILC1, ILC3 and ILC5. The final row shows the expected power in
quadrupole and the average value of S statistics of the $10^6$
Gaussian random statistically isotropic CMB maps. P(S) is the
probability that a random map has a quadrupole-octopole alignment as
high as S.}
\begin{tabular}{|c|c|c|c|}
 \hline
       &  $C_2\,(\mu \rm{K^2})$  &  S  &  P(S)  \\
\hline
 ILC1 & 204.4 &  0.744  &   0.8\%     \\
\hline
 ILC3 & 260.3 &  0.700  &   2.1\%     \\
\hline
 ILC5 & 254.1 &  0.726  &   1.2\%     \\
\hline
 $\Lambda\rm{CDM}$ &  1071.5      &   0.412     &   50.0\%     \\
\hline
\end{tabular}
\end{center}
\end{table}

\begin{figure}
\begin{center}
\includegraphics[height=0.24\textheight,width=0.71\textwidth]{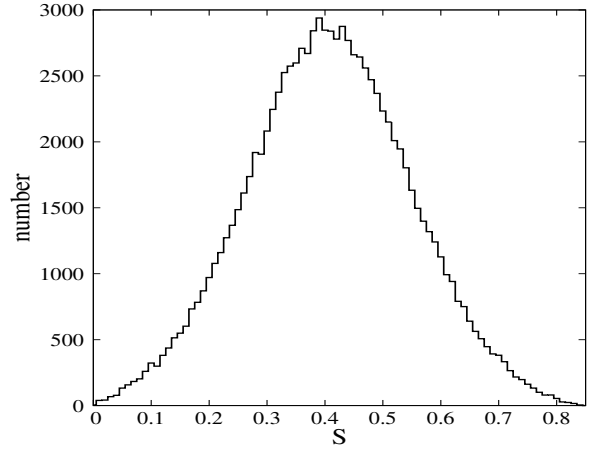}
\caption{Histogram of S statistics for $10^6$ Gaussian random,
statistically isotropic Monte Carlo maps.}
\end{center}
\end{figure}

\begin{figure}
\begin{center}
\includegraphics[height=0.24\textheight,width=0.71\textwidth]{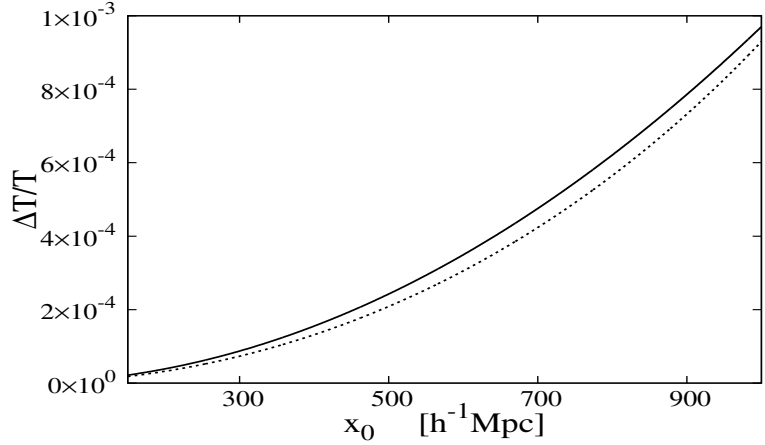}
\caption{The quadrupole amplitudes as a function of distance between
the observer and local gravitational field. The solid line indicates
the SR kinematic quadrupole, and the dotted line represents the
local quadrupole. We assume a higher value for $X_c$, that is
$x_c=1000 h^{-1}\,\rm{Mpc}$ to satisfy $X_0<X_c$.}
\end{center}
\end{figure}

\begin{figure}
\begin{center}
\includegraphics[height=0.24\textheight,width=0.71\textwidth]{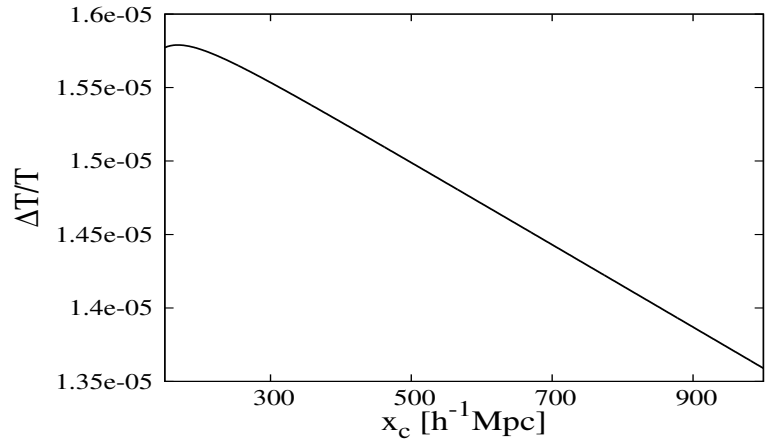}
\caption{The relationship between the amplitudes of local quadrupole
and the radius of the local gravitational field for $x_0=112
h^{-1}\,\rm{Mpc}$.}
\end{center}
\end{figure}

\begin{figure}
\begin{center}
\includegraphics[height=0.24\textheight,width=0.71\textwidth]{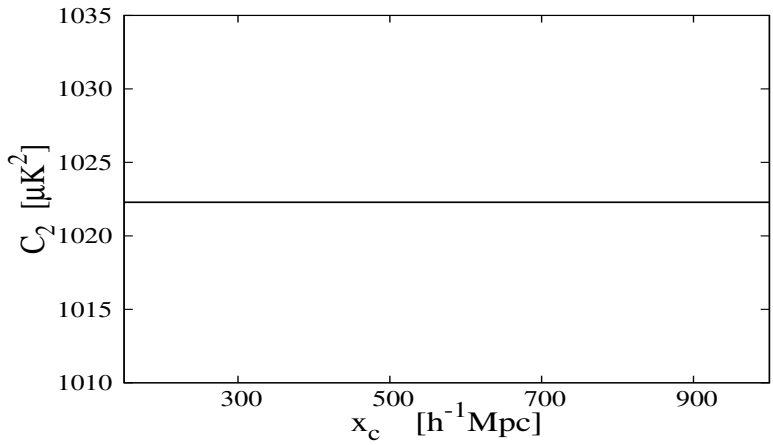}
\caption{Corrected $C_{2}$ for ILC5, when
$x_0=112h^{-1}\,\rm{Mpc}$.}
\end{center}
\end{figure}

\begin{figure}
\begin{center}
\includegraphics[height=0.24\textheight,width=0.71\textwidth]{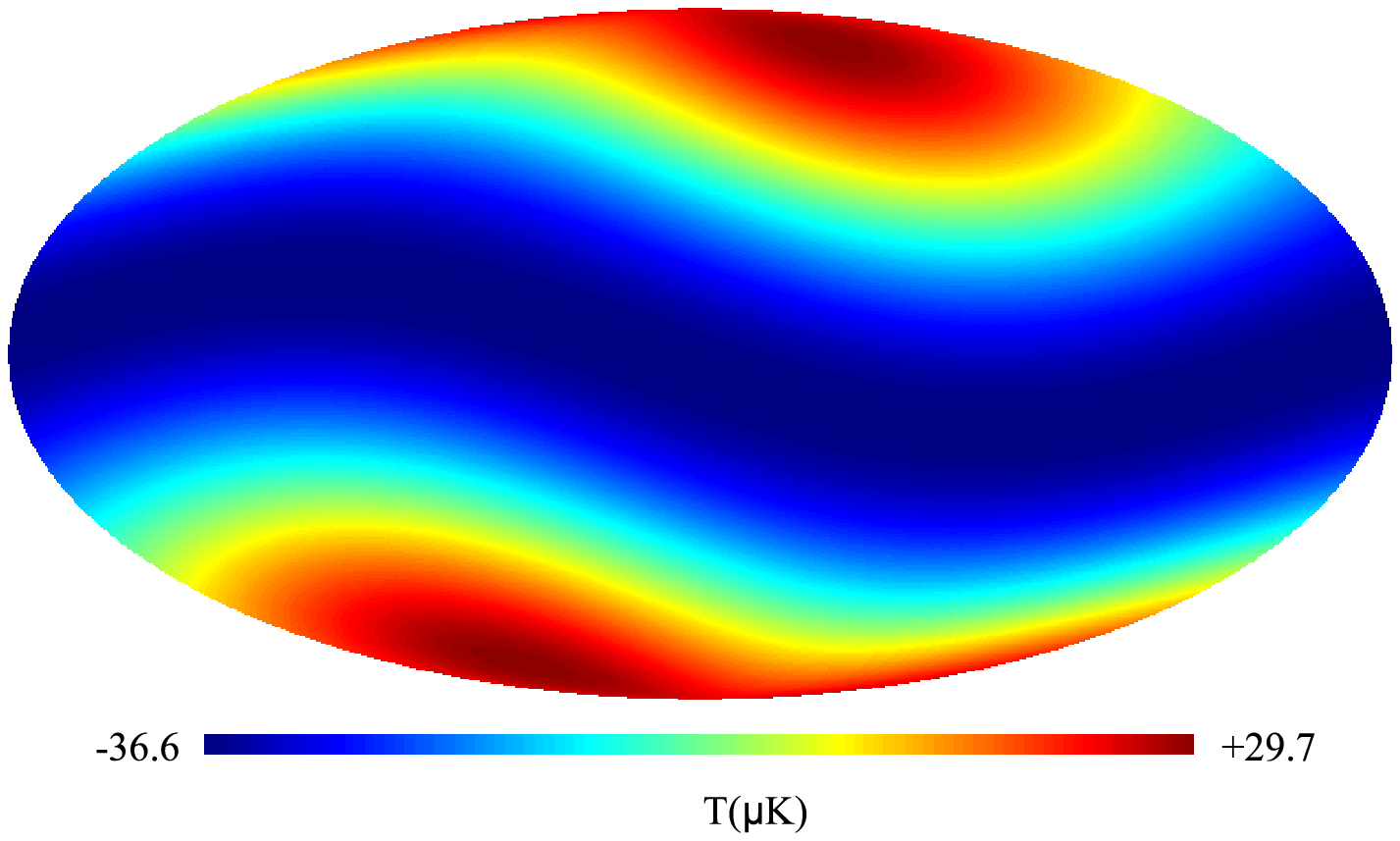}
\caption{The local quadrupole map. The direction of the local
gravitational field is $(284^{\circ},74^{\circ})$, the density
fluctuation is $10^{-3}$, and it is about $112h^{-1}\,\rm{Mpc}$ away
from us.}
\end{center}
\end{figure}

\begin{table}[!hbp]
\begin{center} \caption[]{Power in quadrupole $C_2$ and alignments of
the "foreground-corrected" CMB maps for ILC1, ILC3 and ILC5.}
\begin{tabular}{|c|c|c|c|}
\hline
       &  $C_2\,(\mu \rm{K^2})$  &  S  &  P(S)  \\
\hline
 ILC1-corr &  1064.2   &  0.317   &   75.1\%   \\
\hline
 ILC3-corr &  1034.2   &  0.371   &   61.5\%   \\
\hline
 ILC5-corr &  1022.3   &  0.368   &   62.2\%   \\
\hline
 $\Lambda\rm{CDM}$ &  1071.5   &  0.412      &   50.0\%   \\
\hline
\end{tabular}
\end{center}
\end{table}

\end{document}